\documentclass[fleqn,twoside]{article}
\usepackage{espcrc2}
\usepackage{amsmath}
\usepackage{amssymb}
\usepackage{graphicx}
\usepackage{epsfig}
\usepackage[figuresright]{rotating}

\newcommand{\nue}{\nu_{e}}
\newcommand{\num}{\nu_{\mu}}
\newcommand{\nut}{\nu_{\tau}}

\newcommand{\nuebar}{\overline{\nu}_{e}}
\newcommand{\numbar}{\overline{\nu}_{\mu}}

\newcommand{\sinSqr}{\sin^{2}2\theta_{23}}
\newcommand{\deltaMSqr}{\Delta m^{2}_{23}}

\title{First MINOS Results with the NuMI Beam}

\author{B. J. Rebel\address{Fermilab, PO Box 500, Batavia, IL 60510, USA} for the MINOS Collaboration}
       
\begin{document}

\begin{abstract}
The results from the initial exposure of the MINOS detectors to neutrinos produced by the Fermilab NuMI beam are reported here.  The exposure consisted of $1.27\times 10^{20}$ 120 GeV protons incident on the NuMI target. The data show the observation of 215 neutrinos with energies below 30 GeV while $336\pm14.4$ events were expected.  The data are consistent with $\nu_{\mu}$ disappearance via neutrino oscillations with $|\deltaMSqr|=2.74^{+0.44}_{-0.26}\times 10^{-3}\text{ eV}^{2}/c^{4}$ and $\sinSqr > 0.87$ (68\% C.L.).
\vspace{1pc}
\end{abstract}

\maketitle

There is substantial evidence that $\num$ oscillate into $\nut~$\cite{ref:osc1,ref:osc2,ref:osc5}. The Main Injector Neutrino Oscillation Search (MINOS) was designed to study this hypothesis.   The probability that a $\num$ of energy $E$ remains a $\num$ after traveling a distance $L$ is 
\begin{equation}
P_{\num\rightarrow\nut} = 1 - \sinSqr\sin^{2}(1.27|\deltaMSqr| L/E),
\label{eq:oscprob}
\end{equation}
where $|\deltaMSqr|$ has units of eV$^{2}$, $L$ has units of km and $E$ is in GeV~\cite{ref:parke}.  MINOS tests the oscillation hypothesis by making two measurements of a beam of $\num$ produced in the Neutrinos at the Main Injector (NuMI) beam at Fermilab.  The first measurement occurs at the Near Detector (ND) located onsite at Fermilab, 1~km from the production point; the second measurement is made at the Far Detector (FD) located 735~km away in the Soudan Underground Mine in Soudan, Minnesota, USA.  MINOS extracts the oscillation parameters by comparing the reconstructed energy spectra of the $\num$ at the ND and FD. 

The NuMI beam is produced using 120 GeV protons from the Main Injector.  The protons are delivered in 10~$\mu$s spills, each of which contains up to $3.0\times10^{13}$ protons.  
%The protons extracted from the Main Injector are bent downward by $3.3^{\circ}$ in order to point at the MINOS detectors.  
Two parabolic horns are pulsed with 200~kA of current in order to focus the produced $\pi^{+}$ and $K^{+}$ toward the detectors.  The horns are pulsed in such a way as to maximize the production of neutrinos in the energy range of 1-3~GeV.  A total of $1.27\times10^{20}$ protons on target (POT) were delivered for the data described below.  The neutrino beam is made of 92.9\% $\num$, 5.8\% $\numbar$, 1.2\% $\nue$ and 0.1\% $\nuebar$. 

The MINOS detectors are steel-scintillator tracking calorimeters with toroidal magnetic fields averaging 1.3~T~\cite{ref:minos1}.  The steel planes are 2.54~cm thick and the scintillator is mounted to the steel.  The scintillator planes are made of 4.1~cm wide and 1~cm thick strips.  The strips in each plane are rotated 45$^{\circ}$ from the vertical and the strips in successive planes are rotated 90$^{\circ}$ from each other.  The light produced in the scintillator is collected in 1.2~mm wavelength shifting fibers embedded in the scintillator.  The fibers transport the light to the multi-anode photomultiplier tubes (PMTs).  The detectors were made as similar as possible in order to cancel the majority of the uncertainties in the neutrino interaction modeling and detector response.  Both detectors yield 6-7~photoelectrons per plane for normally incident minimum ionizing particles.  The main design differences between the two detectors are due to the much higher  rate ($\sim10^{5}$ times) in the ND than in the FD.

The FD is 705~m below the surface, has a mass of 5.40~kton, and is composed of 484 instrumented planes.  The detector has a regular octagonal cross section and is 8~m wide.  The scintillator is read out at both ends of the strips using Hamamatsu M16 PMTs.  
%Each pixel of a PMT reads out the fibers from 8 scintillator strips that are spaced about 1~m apart in the plane.  The pattern for coupling the fibers to the PMTs is different on the two sides of the planes to allow resolution of ambiguities in which strip produced an observed signal.  The first step of the data reconstruction in the FD is to remove these ambiguities. 
The front end readout electronics are designed to provide high precision timing information.  The FD data were blinded until the procedures for event selection and energy spectrum prediction were defined and understood.  The blinding procedure hid a substantial and unknown fraction of the events in the FD.% with the precise fraction and energy spectrum of the hidden set unknown.

The ND is 103~m below the surface, has a mass of 0.98~kton, and is composed of 282 planes.  The planes have an irregular octagonal cross section and are 4~m tall and 6~m wide.  The geometry of the planes optimizes containment of hadronic showers and allows for the magnetic field to be similar to that in the FD.  
%It has two logical components, the calorimeter and spectrometer.  In the calorimeter each plane is instrumented while in the spectrometer only every fifth plane is instrumented.  
The front end electronics in the ND are designed to handle the high rate of interactions observed with each Main Injector spill.  Up to 10 interactions can be observed in the ND for each spill of protons.  As such the first step in the reconstruction of the ND data is to use timing and spacial information to separate the individual interactions.  The data in the ND were not blinded.

The energy of each neutrino interaction is found in the same way for both detectors.  Muon tracks are found and their curvature in the magnetic field is fit to determine their energy.  The hadronic showers are also found and their energy is determined.  
%For $\num$ charged current (CC) interactions, the total energy is the sum of the muon and hadronic shower energies.  
The events selected in both detectors were required to have visible energy, $E_{vis}$,  less than 30~GeV and the events had to have a negatively charged track, a requirement chosen to select only $\num$ interactions.  
%The requirement on the charge of the track was to select events from $\num$ interactions and exclude those from $\numbar$ interactions. 
A fiducial volume was defined to contain the hadronic energy of the event and reject background cosmic ray muons.  The events were also required to occur within a 50~$\mu$s window surrounding the spill time.  The background due to cosmic ray muons in the ND is negligible.  However, an additional constraint to reduce the cosmic ray background was imposed in the FD;  the direction of the reconstructed track at its vertex had to be within 53$^{\circ}$ of the neutrino beam direction.  The background due to cosmic ray events in the FD is estimated to be $<0.5$ events (68\% C.L.) as there are no events occurring within the 50~$\mu$s window.  

\begin{figure}
\centerline{\epsfig{file=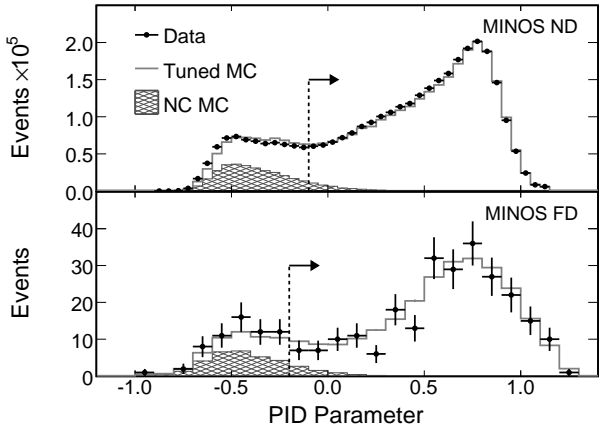, width=3.5in}}
\caption{Data and Monte Carlo distributions for the PID parameter in the ND (top) and FD (bottom).  The dotted vertical lines show the position of the cut for CC-like events and the arrows indicate the selected events.}
\label{fig:pid}
\end{figure}

As the MINOS detectors record both neutral current (NC) and CC interactions, a particle identification parameter (PID) was defined to determine whether an event was NC or $\num$ CC.  The PID incorporated the probability density functions for the event length, fraction of energy contained in the track and average track pulse height per plane to separate the events into these two categories.  Figure~\ref{fig:pid} shows the PID for the ND and FD data overlaid with the Monte Carlo NC and CC distributions.  The events with PID $> -0.2$ were categorized as CC events in the FD; in the ND the requirement was PID $> -0.1$.  The values of the PID were chosen so that the purity of the samples in the ND and FD were both about 98\%.  The efficiencies for selecting $\num$ CC events with $E_{vis} < 30$~GeV in the fiducial volume are 74\% for the FD and 67\% for the ND.  

The measured energy spectrum in the ND is used to predict the unoscillated spectrum in the FD.
%Pion decay kinematics and beamline geometry can introduce shape differences in the ND and FD spectra of up to $\sim20$\% on either side of the peak that are unrelated to oscillations.  The FD spectrum prediction must take these differences into account.  
The method used by MINOS to predict the FD spectrum is the {\it Beam Matrix} method~\cite{ref:adam}. In this method the ND data are used to measure effects such as beam modeling, neutrino interactions and detector response that are common to both detectors.  The beam simulation is used to derive a transfer matrix that relates $\num$ in the two detectors via their parent hadrons.  The matrix element $M_{ij}$ gives the relative probability that the distribution of secondary hadrons producing the observed $\num$ of energy $E_{i}$ in the ND will produce the observed $\num$ of energy $E_{j}$ in the FD.  The reconstructed energy spectrum in the ND is translated into a flux by correcting the simulated ND acceptance and then dividing by the calculated cross-sections for each energy bin.  The flux is multiplied by the matrix to yield the predicted unoscillated FD flux.  The final step in the process is to do the inverse correction for cross-section and FD acceptance, resulting in the predicted visible energy spectrum.  As a cross check of this method, the prediction was also done with the {\it ND Fit} method, which minimizes differences between the ND data and Monte Carlo by modifying the parameters associated with neutrino interactions and detector response.  The FD Monte Carlo is adjusted by using the best-fit values of those parameters.

\begin{figure}
\centerline{\epsfig{file=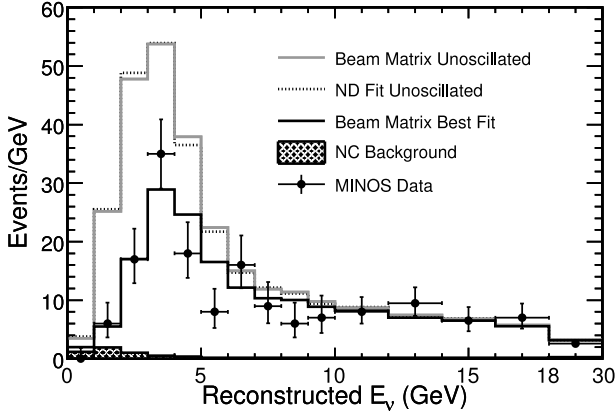, width=3.5in}}
\caption{Comparison of the FD spectrum with predictions for no oscillations for both analysis methods.  Also shown are the best fit hypothesis from the {\it Beam Matrix} method and the estimated NC background.  The last bin contains events with $18<E_{vis}<30$~GeV.}
\label{fig:spectrum}
\end{figure}

The FD data set contains a total of 215 events with $E_{vis}< 30$~GeV compared to the unoscillated expectation of $336.0\pm14.4$.  The uncertainty is due to systematic uncertainties associated with (a) the fiducial mass calculation and POT counting accuracy (4\%), (b) the hadronic energy scale (11\%) and (c) the NC component (50\%).  
%For $E_{visible}< 10$~GeV, there are 122 events with an expectation of $238.7\pm10.7$.  
The observed energy spectrum is shown in Fig.~\ref{fig:spectrum} along with the predicted spectra from both methods.  

\begin{figure}
\centerline{\epsfig{file=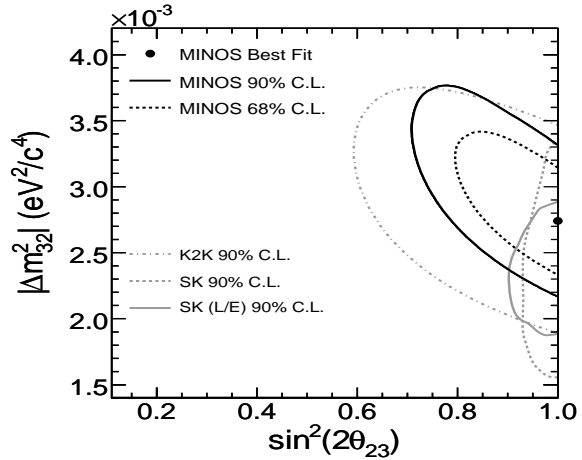, width=3.2in, height=2.5in}}
\caption{Confidence intervals for the best fit using the {\it Beam Matrix} method.  The intervals include the effects of the systematic uncertainties.  Also shown are the contours from the previous highest precision experiments~\cite{{ref:osc1,ref:osc2,ref:osc5}}.}
\label{fig:contours}
\end{figure}

The data were fit to the hypothesis of $\num\rightarrow\nut$ oscillations, and the 
% by minimizing the $\chi^{2}$ statistic, 
%\begin{equation}
%\chi^{2} = 2\sum_{i=1}^{15}(e_{i}-o_{i}+o_{i}\log\frac{o_{i}}{e_{i}}) + \sum_{j=1}^{3}\frac{\Delta s_{j}^{2}}{\sigma^{2}_{j}}.
%\label{eq:chisqr}
%\end{equation}
%In eq.(\ref{eq:chisqr}) the first sum is over the number of bins in the energy spectrum, with $o_{i}$ the observed number of events in bin $i$ and $e_{i}$ are the expected number of events due to oscillations including events from $\nut$ produced in the oscillation process.  The second sum is over the parameters describing the systematic uncertainties.  The $s_{j}$ are allowed to vary in the fit but are constrained by the size of the uncertainty on the parameters discussed above.  
fit has $\sinSqr>0.87$ at the 68\% C.L. The best fit for the mass squared difference is $|\deltaMSqr| = 2.74^{+0.44}_{-0.26}\times 10^{-3}\text{eV}^{2}/c^{4}$ with the fit probability of 8.9\%.  The 68\% and 90\% confidence intervals are shown in Fig.~\ref{fig:contours} along with the 90\% contours from the previous highest precision experiments.  

\end{document}